%% file: celeste_ipdps.tex
\documentclass[10pt, conference, compsocconf]{IEEEtran}

\usepackage{cite}
\usepackage{color}
\usepackage{xspace}
\usepackage[pdftex]{graphicx}
\usepackage{subcaption}
\graphicspath{{figures/}}
\DeclareGraphicsExtensions{.pdf,.jpeg,.png}

\usepackage[cmex10]{amsmath}
\usepackage{amssymb}
\usepackage{url}

\newcommand{\partCPUtm}{Intel\textsuperscript{\textregistered} Xeon\textsuperscript{\textregistered}\xspace}

\begin{document}

\title{Learning an Astronomical Catalog of the Visible Universe through\\Scalable Bayesian Inference}

\author{\IEEEauthorblockN{
Jeffrey Regier\IEEEauthorrefmark{1},
Kiran Pamnany\IEEEauthorrefmark{2},
Ryan Giordano\IEEEauthorrefmark{3},
Rollin Thomas\IEEEauthorrefmark{5},
David Schlegel\IEEEauthorrefmark{4},
Jon McAuliffe\IEEEauthorrefmark{3} and
Prabhat\IEEEauthorrefmark{5}}
\IEEEauthorblockA{\IEEEauthorrefmark{1}Department of Electrical Engineering and Computer Science, University of California, Berkeley}
\IEEEauthorblockA{\IEEEauthorrefmark{2}Parallel Computing Lab, Intel Corporation}
\IEEEauthorblockA{\IEEEauthorrefmark{3}Department of Statistics, University of California, Berkeley}
\IEEEauthorblockA{\IEEEauthorrefmark{4}Physics Division, Lawrence Berkeley National Laboratory}
\IEEEauthorblockA{\IEEEauthorrefmark{5}NERSC, Lawrence Berkeley National Laboratory}
}

\maketitle
\thispagestyle{plain}
\pagestyle{plain}

\begin{abstract}
%DO NOT USE SPECIAL CHARACTERS, SYMBOLS, OR MATH IN YOUR TITLE OR ABSTRACT.
Celeste is a procedure for inferring astronomical catalogs that attains
state-of-the-art scientific results. To date, Celeste has been
scaled to at most hundreds of megabytes of astronomical images: Bayesian posterior
inference is notoriously demanding computationally. In this paper, we report on 
a scalable, parallel version of Celeste, suitable for learning catalogs from
modern large-scale astronomical datasets. Our algorithmic innovations include
a fast numerical optimization routine for Bayesian posterior inference and a
statistically efficient scheme for decomposing astronomical optimization problems
into subproblems.

Our scalable implementation is written entirely in Julia, a new high-level dynamic programming
language designed for scientific and numerical computing.  We use Julia's high-level
constructs for shared and distributed memory parallelism, and demonstrate
effective load balancing and efficient scaling on up to 8192 Xeon cores on
the NERSC Cori supercomputer.
\end{abstract}

%Celeste is written entirely in Julia, a new programming language that
%provides high-level syntax like Python and MATLAB, and computational efficiency
%like C/C++ and Fortran. We use Cray MPI-3 and Julia's threading for distributed
%parallelism and dynamic load balancing across 52,000 Xeon cores on the NERSC
%Cori platform.  \end{abstract}
% kpamnany: expand first sentence -- inferring astronomical catalogs of
% celestial objects from images? state-of-the-art scientific results -- of what
% sort? accuracy? and performance? also, we won't be using all 55 TB. How big is
% a 16x16 RA Dec box?


% \fi
\IEEEpeerreviewmaketitle

\input{introduction.tex}

\input{related_work.tex}

\section{Celeste}

Celeste approximates the posterior distribution
using a variational inference procedure that facilitates parallel processing.
The Celeste model and the variational objective function are
described in earlier work about a single-threaded version of Celeste~\cite{regier2015celeste}.
Our approach to optimizing the objective function, based on a trust-region Newton's
method, is new, as is the approach to parallelism.

\subsection{The Celeste Model}\label{model}

\begin{figure}[!t]
\centering
\includegraphics[width=2in]{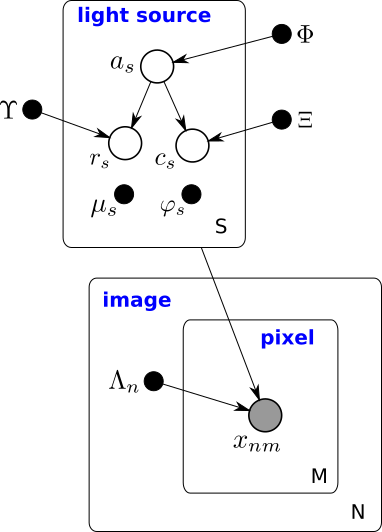}
\caption{The Celeste graphical model. Shaded vertices represent observed random
variables. Empty vertices represent latent random variables.  Black dots
represent constants. Constants denoted by uppercase Greek characters are fixed
throughout our procedure. Constants denoted by lowercase Greek
letters are inferred, along with the posterior distribution of 
the latent random variables. Edges signify permitted conditional
dependencies. Plates (the boxes) represent independent replication.}

\label{graphical_model}
\end{figure}

A generative Bayesian model is a joint probability distribution over observed random
variables (the pixel intensities) and unobserved, or latent, random
variables (the catalog entries).  The Celeste model is represented graphically
in Figure~\ref{graphical_model}.

\subsubsection*{Images} Each of $N$ images has fixed metadata $\Lambda_n$,
describing its sky location and the atmospheric conditions at the time of the
exposure.  Each image contains $M$ pixels.  Each pixel intensity is an observed
random variable, denoted $x_{nm}$, that follows a Poisson distribution with a
rate parameter $F_{nm}$ unique to that pixel.  $F_{nm}$ is a deterministic
function of the catalog (which includes random quantities) and the image
metadata.  It follows that pixel intensities are conditionally independent
given the catalog.

\subsubsection*{Light sources} For each of $S$ light sources, a Bernoulli
random variable $a_s$ indicates whether it is a star or a galaxy; a lognormal
random variable $r_s$ denotes its brightness in a particular band of emissions
(the ``reference band''); and a multivariate normal random vector represents
its colors---defined formally as the log ratio of brightnesses in adjacent
bands, but corresponding to the colloquial definition of color. These random
quantities have prior distributions with parameters $\Phi$, $\Upsilon$, and
$\Xi$, respectively. These parameters are learned from pre-existing
astronomical catalogs.  Additionally, each light source is characterized by
(non-random) vectors $\mu_s$ and $\varphi_s$. The former indicates the light
source's location and the latter, if the light source is a galaxy, represents
its shape, scale, and morphology. Though non-random, these vectors are
nonetheless learned within our inference procedure.

It is straightforward to sample collections of ``synthetic'' astronomical images
from the Celeste model, given the preceding description.
Indeed, we do generate data in this way for testing purposes.
Our primary use for the model, however, is to compute the distribution of its
unobserved random variables conditional on a particular collection
of (real) astronomical images. This distribution is known as the posterior.
Exact posterior inference is computationally intractable for the Celeste
model, as it is for most non-trivial Bayesian models.
Instead, we use variational inference to approximate the posterior.

\subsection{Variational Inference}\label{vi}

Let $x := \{x_{nm}\}_{n=1,m=1}^{N,M}$ denote all pixels.
Let $z := \{a_s, r_s, c_s\}_{s=1}^S$ denote all the latent random variables.
The posterior distribution on $z$, i.e. $p(z|x)$, combines our prior knowledge
with the new information contained in the data.

Variational inference (VI) chooses a distribution $q$ from a class of
candidates to approximate the posterior $p(z|x)$ by maximizing the following
lower bound on the log probability of the data~\cite{blei2016variational}:
\begin{align}
\log p(x_{11},\ldots,x_{NM})
&\ge \mathbb{E}_q \left[ \log p(x, z) - \log q(z) \right] \label{elbo}\\
&=: \mathcal L(\theta).
\end{align}
This lower bound holds for every $q$, which follows from Jensen's inequality.

We take the density $q$ to factorize across light sources and across each 
light source's latent variables:
\begin{align}
q(z) = \prod_{s=1}^S q(r_s | a_s)q(c_s | a_s)q(a_s).
\end{align}
In our formulation, each factor of $q$ is an exponential family that is conjugate to the corresponding prior distribution: $q(r_s | a_s)$ is univariate lognormal, $q(c_s | a_s)$ is multivariate normal, and $q(a_s)$ is Bernoulli.

We index the class of candidate $q$'s by the real-valued vector
$\theta$. Then, maximizing over $q$ is equivalent to maximizing over $\theta$.
It can be shown that the maximizer of $\mathcal L$
also minimizes
\begin{align}
D_{\mathrm{KL}} \left( q(z), p(z | x) \right),
\end{align}
the Kullback-Leibler divergence to the 
posterior~\cite{blei2016variational}. The objective function $\mathcal L$ may also be maximized 
over the model
parameters $\{\mu_s, \varphi_s\}_{s=1}^S$, simultaneously, to estimate them.

The posterior distribution typically will not have the form of
the candidate variational distributions, so the minimum Kullback-Leibler divergence
will not be zero, and the posterior distribution will not be recovered
exactly. For classes of variational distributions that factorize, such as ours,
the variational distribution that minimizes Kullback-Leibler divergence tends to
underestimate the posterior's variance. Techniques exist to correct this bias~\cite{giordano2015linear}.

Though $q$ factorizes across the light sources, the light
sources are still coupled in $\mathcal L$ by $p(x,z)$: pixels may receive photons from multiple light
sources. The parameter vector $\theta$ contains 32 entries per light source,
for each of hundreds of millions of light sources.

To simplify optimization, rather than maximizing $\mathcal L$ with respect to
all of $\theta$ jointly, we optimize subsets of $\theta$ corresponding to the
parameters for distinct light sources independently, while holding the
parameters for other light sources fixed at estimates from previous
astronomical surveys.  For light sources that do not overlap with any other
light sources, the optimal point returned by this technique is no different
than had we optimized $\theta$ jointly across all light sources. For light
sources that do overlap with others, we anticipate that existing estimates of
neighboring light sources will be sufficiently accurate. Regardless,
$\max_\theta \mathcal L$ is a lower bound on the log probability of the data.
Our alternative optimization procedure may find a looser lower bound, but with
significant computational savings from decoupling optimization of the light
sources.

%{\textcolor{blue}{At some point, perhaps here, we should also mention that we have to approximate intractable
%integrals within $\mathcal{L}$.}}

In~\cite{regier2015celeste}, we used L-BFGS~\cite{wright1999numerical} to optimize $\mathcal L$.
However, some light sources require thousands
of L-BFGS iterations to converge. These light sources dominate runtime.
The approach based on L-BFGS is too inefficient for large-scale optimization.

For our scalable version of Celeste, we use Newton's method, with updates
constrained by a trust region, to optimize $\mathcal L$
for each source. Newton's method consistently reaches
machine tolerance within 50 iterations for our optimization problems.
Furthermore, to maximize efficiency, rather than using automatic differentiation,
we manually compute the gradient and the (dense) Hessian for each
light source. 

\subsection{Parallel Work Decomposition}\label{parallelism}

To find the optimal parameters for a light source, Celeste must
load the following: (a) all the images containing that light source,
(b) an initial estimate of its parameters, and (c) initial estimates of any
nearby light source that may overlap with it.

Image data is at least an order of magnitude larger than the initial parameter
estimates. An image is stored as a collection of five files that are
roughly 60 MB in aggregate. Approximately 500 light sources appear in each
image. We aim to limit the I/O cost associated with loading images repeatedly.

On a fast, modern processor, Celeste finds the optimal parameters for a
light source in anywhere from one second to over two minutes, with most
sources taking less than five seconds. Because the work is irregular, we
use dynamic scheduling to balance load.

These two aims present a challenge. We must limit I/O---multiple 
processes loading the same image repeatedly, once for each light
source in in it, would lead to a severe I/O bottleneck. Yet we cannot
distribute work at the image level, as that could lead to load imbalance
on the order of minutes.

We considered two work decomposition strategies.
The first strategy partitions
the sky into equal-size contiguous regions smaller than an image.
Each region corresponds to a task. If an image contains $R$ such regions,
then at most $R$ nodes must load the same image, limiting the I/O burden.
However, our experiments with this approach still showed high load
imbalance. Although cosmological theory predicts a certain type of
uniformity of light sources across the Universe, in practice we find that
some regions of the sky have many sources while other regions have few to
none.

For our second strategy, candidate light sources correspond to tasks. Given
an existing catalog of light sources, we dynamically schedule batches of
light sources to nodes for optimization. Since each image typically
contains many light sources, this second strategy could potentially require
the same image to be loaded many times by different processes. We use
two techniques to reduce the I/O burden: first, we load all images from
disk into the memory of all the participating processes, using a global
array implementation, thus converting a slow, disk-bound operation into a much
faster one-sided RMA operation on a high-performance interconnect fabric. Second, we
use a task scheduling scheme that distributes tasks in spatially aware batches, thereby
reducing the frequency of multiple processes requiring the same image
data.

We implement the second strategy for our experiments.

\subsection{Implementation}

The current implementation proceeds in three phases:

\subsubsection*{1. Load images} All required images are loaded into a global
array. Each process loads images concurrently, using only a single thread
since this stage is I/O bound.

\subsubsection*{2. Load catalog} An existing catalog of candidate light
sources is loaded into a second global array. Each entry includes initial
estimates of the suspected light sources' parameters (recall that these
initial estimates are used for rendering neighboring light sources during
optimization). The candidate light sources are ordered according to their
spatial position, thus nearby light sources are also close together in the 
global array. This ordering reduces communication but is not necessary for
correctness.

\subsubsection*{3. Optimize sources} Each entry in the catalog global array
is a task. We use the Dtree scheduler~\cite{pamnany2015dtree} to distribute
batches of contiguous indices into the catalog global array, to processes.
The batches are small to help balance load.

A process uses multiple threads to optimize the light source in the region
it is assigned. These threads share a process-level cache of
images and catalog entries. Each thread retrieves the next index from the
batch assigned to the process, fetches that entry from the catalog global
array, and then checks the cache for the necessary images and neighbor
catalog entries, fetching them from the global arrays as needed.

\subsection{Language considerations}

Celeste is implemented entirely in Julia, a high level dynamic programming language.
Julia offers many desirable features for scientists and engineers---familiar
syntax, interactive development, a high level of abstraction, garbage
collection, dynamic types, and excellent integrated open-source libraries
for many areas, including linear algebra and signal processing. Julia's compiler is based on
LLVM. It uses type inference, JIT compilation, vectorization, and inlining
to achieve excellent performance, competitive with statically-typed languages
such as C/C++.

This combination of features---expressiveness and ease of use, together with
high performance---makes Julia a compelling choice for scientific and
numerical computing. While still at an early stage with version $0.5$ (as of October 2016), the Julia language is maturing rapidly. As such, using Julia presents advantages as well as challenges, particularly relating to parallelism.
We discuss these challenges in Section~\ref{julia}.

\subsection{Multi-process scaling}

Julia offers distributed parallelism capabilities in its base libraries.
This functionality is built around remote procedure calls (RPC), an
abstraction that has limited application to HPC applications, which tend to
be mostly data parallel, and have stringent latency and bandwidth needs.

Julia applications can directly use MPI via a wrapper package. However,
MPI does not offer high-level abstractions that are easy to use and
expressive. The PGAS model~\cite{pgas}, on the other hand, was developed
precisely to improve the productivity of distributed parallel programming.

Thus, we have developed a lean and fast global arrays library that
implements essential parts of the Global Arrays
interface~\cite{nieplocha2006advances}. We use MPI-3 as the transport
layer for our library; get and put operations on global array
elements make use of one-sided RMA operations that are supported in
hardware on most supercomputer fabrics~\cite{hoefler2015rma}. We have
also developed a Julia wrapper package for this library, making
the PGAS programming model available to Julia applications such as
Celeste.

\subsection{Distributed dynamic scheduling}

To serve the dynamic scheduling needs of Celeste, we have written a
Julia wrapper for Dtree~\cite{pamnany2015dtree}, a distributed dynamic
scheduler that has been shown to be capable of balancing load for a
diverse range of irregular tasks, even at petascale. The effectiveness
of this scheduler can be seen in the minimal scheduling overhead we
observed in our experiments.

Dtree organizes processes into a short tree for task distribution; the
tree fan-out is configurable and allows for multiple parents in order
to eliminate any bottleneck arising from too many children sharing a
single parent. Given a total number of tasks, $T$, parents in the tree
distribute batches of number ranges, $f$-$l$, where $f, l <= T$ in
response to requests from child processes. The size of each batch,
$n = l-f$, reduces as $T$ is approached; this balances load.

Celeste uses the size of the catalog global array, that is, the
number of candidate light sources, as the total number of tasks, $T$,
as previously described.

\input{sdss}

\section{Hardware Platform}
%Edison is a Cray XC30, with a peak performance of 2.57 petaflops/sec, 133,824
%compute cores, 357 terabytes of memory

%Celeste utilized the Cori Phase~I system at the National Energy
Our experiments are conducted on the Cori supercomputer at the National
Energy Research Scientific Computing Center (NERSC). Cori is a Cray XC40
supercomputer. We used Cori Phase~I (also known as the ``Cori Data
Partition'') which has 1,630 compute nodes, each containing two \partCPUtm
E5-2698v3 processors\footnote{Intel, Xeon, and Intel Xeon Phi are
trademarks of Intel Corporation in the U.S. and/or other countries.} (16
cores each) running at 2.3~GHz, and 128~GB of DDR4 memory.
%Each core has its own L1 and L2 caches, with 64 KB
%(32 KB instruction cache, 32 KB data) and 256 KB, respectively; there is also a
%40-MB shared L3 cache per socket.
Nodes are linked through a Cray Aries high speed ``dragonfly'' topology interconnect. Datasets used in these experiments were staged on Cori's 30 PB Lustre file system, which has an aggregate bandwidth exceeding 700 GB/s.

\input{scaling_results}

\input{scientific_results}

\input{julia}

\input{conclusions}

\section*{Acknowledgments}
This research used resources of the National Energy Research Scientific
Computing Center (NERSC), a DOE Office of Science User Facility supported by
the Office of Science of the U.S. Department of Energy under Contract No.
DE-AC02-05CH11231.  The authors express their gratitude to Tina Declerck, Doug
Jacobsen, David Paul and Zhengi Zhao who helped make the results presented in
this paper possible. Debbie Bard and Dustin Lang provided expert advice on
astronomy datasets and image processing issues. We are grateful for the
tremendous assistance rendered by various members of the Julia developer
community: Tony Kelman, Jameson Nash, Andreas Noack, Alan Edelman and Viral
Shah addressed various functionality and performance issues in Julia.
Finally, we thank Andy Miller, Ryan Adams and Steve Howard,
who assisted in developing the Celeste codebase. 

Funding for the Sloan Digital Sky Survey IV has been provided by
the Alfred P. Sloan Foundation, the U.S. Department of Energy Office of
Science, and the Participating Institutions. SDSS-IV acknowledges
support and resources from the Center for High-Performance Computing at
the University of Utah. The SDSS web site is www.sdss.org.

\bibliographystyle{IEEEtran}
\bibliography{IEEEabrv,references}

\end{document}

%% file: introduction.tex
\section{Introduction}

The principal product of an astronomical imaging survey is a catalog of celestial objects, such as stars and galaxies. These catalogs are generated by
identifying light sources in survey images and characterizing each according to
physical parameters such as brightness, color, and morphology. Astronomical
catalogs are the starting point for many scientific analyses, such as
theoretical modeling of individual light sources, modeling groups of similar 
light sources, or modeling the spatial distribution of galaxies.
Catalogs also inform the design and operation of subsequent surveys using more
advanced or specialized instrumentation (e.g., spectrographs).  Astronomical
catalogs are key tools for studying the life-cycles of stars and
galaxies as well as the origin and evolution of the Universe itself.

%Astronomical imaging surveys are designed to maximize the area of sky captured
%per night with minimum exposure times that still permit sensitivity to dim
%sources. Large survey cameras have focal planes including a dozen or several
%dozen charge-coupled devices and an aggregate pixel count in excess of 100
%million. A single 100 megapixel (MP) exposure corresponds to 200 MB of raw
%data, depending on the number of bits used to represent a value per pixel.
%Over the lifetime of a multi-year survey many TB of data may be gathered. Data
%generated as the raw data is processed and calibrated further increases the
%storage footprint of the survey imaging data. Among other factors, both the
%data volume and the requirement that catalog representations of individual
%sources be statistically robust make this a challenging problem.

%The Large Synoptic Survey Telescope (LSST)~\cite{lsst} is the next grand
%experiment in imaging survey astronomy. Currently under construction in the
%Chilean Andes, LSST will image the entire visible sky very few nights starting
%in the early 2020's. Attached to LSST will be the largest digital camera
%constructed to date, and each of its 3.2 gigapixel images will cover an area
%of sky nearly equivalent to 18 full Moons. In a single night LSST will
%generate 15 terabytes of raw imaging and over its 10-year planned lifetime the
%survey will collect over 60 petabytes of images of tens of billions or stars
%and galaxies.

Modern astronomical surveys produce vast amounts of complex data. 
The Large Synoptic Survey Telescope (LSST) is slated to capture more than 
15 terabytes of new images on a daily basis~\cite{LSST}; overall the 
instrument will produce 10s--100s of PBs over the lifetime of the project.
Constructing a catalog is computationally demanding for
any method because of the scale of the data. Approaches to date are largely
based on computationally efficient heuristics~\cite{bertin1996sextractor,lupton2005sdss}.
Constructing a high-quality catalog based on a rigorous statistical
treatment of the problem is computationally challenging even for datasets of
modest size.

Bayesian posterior inference is effective for learning astronomical catalogs
because Bayesian models can combine prior knowledge of astrophysics with new data from
surveys in a statistically efficient manner. Bayesian inference also yields accurate
estimates of uncertainty. Because most light sources will be near the detection
limit, uncertainty estimates are as important as the parameter estimates
themselves for many analyses. For example, they enable robust population-level
analysis even when every individual light source is highly 
ambiguous~\cite{schneider2015hierarchical}.

Celeste~\cite{regier2015celeste} is a procedure for inferring an astronomical
catalog from a collection of images, based on a principled statistical model,
that attains state-of-the-art results. While exact posterior inference is
computationally intractable, the approximation proposed in~\cite{regier2015celeste} is nearly
linear time in all relevant quantities: the number of light sources, the number
of pixels, and the number of parameters. Nonetheless, it is a computationally
expensive procedure, never before scaled beyond several hundreds of MBs
of imaging data. The procedure described in~\cite{regier2015celeste} is
single-threaded.

In this paper, we present a new, parallel version of Celeste that enables
scaling the procedure to large datasets on large clusters. Our contributions include:
\begin{itemize}
\item A fast numerical optimization routine for Bayesian posterior inference.
\item A statistically efficient scheme for decomposing astronomical
optimization problems into sub-problems.
%\item \textcolor{magenta}{Something about our parallelism strategy?}
%\item \textcolor{magenta}{Something about reporting strong and weak scaling results?}
\item A demonstration of the viability of using the high-level Julia programming language to implement a complex, real-world data analytics application on a contemporary HPC platform.
\item Development and optimization of a Julia package to enable multi-node distributed memory parallelism. 
\end{itemize}

%Celeste is written entirely in Julia~\cite{bezanson2014julia}, a high-level
%dynamic programming language designed for technical computing. Julia provides
%ease and expressiveness for high-level numerical computing like R, MATLAB,
%and Python, but also offers good performance similar to C/C++ and Fortran by
%using type inference and LLVM-based just-in-time (JIT) compilation.

%% file: related_work.tex
\section{Related Work}

To date, astronomical catalogs have been constructed primarily by
heuristics---algorithms that may be intuitively reasonable, but that do not
follow from a statistical model of the data. 
For the dataset we analyze in this work, discussed in Section~\ref{sdss}, 
the state-of-the-art software pipeline is a heuristic called 
``Photo''~\cite{lupton2005sdss}.
Heuristics have a number of
shortcomings. First, they typically cannot make effective use of prior
information, because it is unclear how to ``weight'' it in relation to new
information from the survey.  Yet much is known about stars' and galaxies'
colors, brightness, and shapes before new data is collected---both from
previous surveys and from astrophysical theory.

Second, heuristics do not effectively combine knowledge from multiple image
surveys, or even from multiple overlapping images from the same survey. They
may ``co-add'' the overlapping images, but this effectively discards properties unique to
each image, like the atmospheric conditions at the time of exposure, or the
exact alignment of the pixels.

Third, heuristics do not correctly quantify uncertainty of their estimates.
They may flag some estimates as particularly unreliable. But confidence
intervals follow only from a statistical model.  Without modeling the arrival
of photons as a Poisson process, for example, there is little basis for
reasoning about uncertainty in the underlying brightness of objects.

These shortcomings are addressed by the Bayesian formalism.  Unknown
quantities, such as the catalog entries for our problem, are modeled as
unobserved random variables with prior distributions.  Then the posterior
distribution, that is, the conditional distribution of the latent variables
given the data, encapsulates knowledge about the latent variables.
Unfortunately, exact Bayesian posterior inference is computationally
intractable for all but the simplest statistical models.

In Tractor ~\cite{thetractor}, rather than inferring the posterior, the mode of
the posterior is learned through maximum a posterior (MAP) estimation.
Though MAP estimation is scalable, and retains many of the advantages of
Bayesian modeling, it does not provide uncertainties,
as one would obtain from posterior inference.

Approximate posterior inference is an alternative to MAP estimation and an alternative to heuristics.
Markov Chain Monte Carlo (MCMC) is the most common approach to approximate posterior inference.
At each iteration, MCMC draws a sample from an approximation to 
the posterior distribution.
Unfortunately, consecutive samples are not independent.
MCMC can require tens of thousands of samples to approximate the
posterior. Thus, to date MCMC has only been used to infer properties of
small collections of stellar and galactic images relative to the sizes 
of modern astronomical datasets~\cite{brewer2013probabilistic}.
Moreover, even given tens of thousands of MCMC samples, it is generally
unknown whether enough samples have been collected to adequately represent
the posterior---the number of samples needed could, for example, grow
exponentially in some dimension of the model.

Variational inference (VI) is an alternative to MCMC that uses numerical
optimization to find a distribution that approximates the posterior without
sampling~\cite{blei2016variational}. In practice, the resulting optimization
problem is often orders of magnitude faster to solve compared to MCMC approaches. 

Scaling VI to large astronomical datasets is nonetheless challenging.  The
largest published applications of VI have been to text mining, where topic
models are fit to several gigabytes of text~\cite{hoffman2013stochastic}. The
SDSS dataset is four orders of magnitude larger than that. Moreover, most
models learned by variational inference to date have a form that allows
optimization by coordinate ascent, where each update can be computed without
explicitly forming gradients.  The Celeste model does not have this property.
Our new scalable version of Celeste instead bases its optimization on a variant of Newton's
method, with manually computed gradients and Hessians---a considerably more
involved undertaking.

\begin{figure}[!t]
\centering
\includegraphics[width=3in]{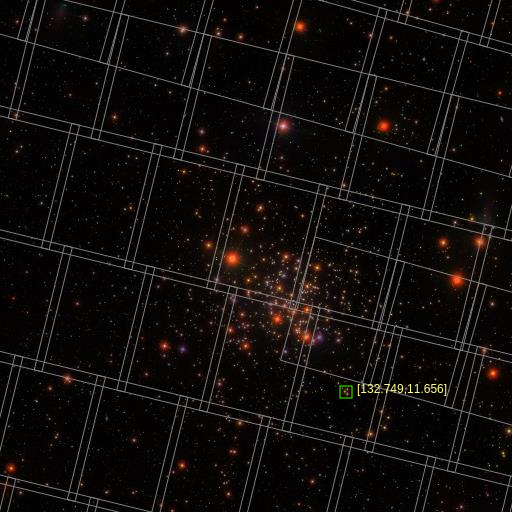}
\caption{SDSS image boundaries. Some images overlap substantially. Some
light sources appear in multiple images that do not overlap. Celeste
uses all relevant data to locate and characterize each light source
whereas heuristics typically ignore all but one image in regions with overlap.
credit:~SDSS DR10 Sky Navigate Tool.}
\label{sdss_fields}
\end{figure}

Furthermore, the datasets for topic modeling, and for many machine learning
tasks, are modeled as $N$ conditionally independent observations given a modest
number of global parameters. Astronomical images, on the other hand, overlap.
Light sources are often imaged many times: Figure~\ref{sdss_fields} shows image boundaries for SDSS data. Images have substantial overlap, and even
non-overlapping parts of different images receive light from a single light
source.  The Celeste statistical model accounts for all of these unique  characteristics.

%% file: sdss.tex
\section{SDSS dataset}\label{sdss}
The Sloan Digital Sky Survey (SDSS)~\cite{stoughton2002sloan} is the
archetypal modern astronomical imaging survey. SDSS Data
Release 12 (DR12) contains almost half a billion individual sources.
It covers 14,555 square degrees of the night sky.
The images are composed of 938,046 ``fields''.
Each field has images of it stored in five different files,
one per filter band.
Each file stores one image in the FITS file format.
The images are $1361\times2048$ pixels, and roughly 12 MB each.
The dataset in total is 55 TB; we processed 250 GB during the
tests reported in this paper.

%% file: scaling_results.tex
\section{Scaling Results}\label{scaling-results}

We ran a variety of experiments to explore the scalability of Celeste using
the parallelization strategy described in Section~\ref{parallelism}.

We analyze Celeste's performance by partitioning the measured runtime for
a job into a number of components: (a) garbage collection time, (b) image load time, (c) load
imbalance, (d) the time taken in retrieving elements of the global arrays
used, (e) dynamic scheduling overhead, and (f) source optimization time.
This partitioning is not precise as it is averaged across nodes and certain
components of runtime are not completely independent. Nonetheless, it is indicative.

We use a \emph{light-sources-per-second} metric to detail source
optimization performance in each scaling test.

\let\thefootnote\relax\footnote{\hrule{}
\vspace{8px}
\noindent{}Software and workloads used in performance tests may have been optimized for performance only on Intel microprocessors. Performance tests, such as SYSmark and MobileMark, are measured using specific computer systems, components, software, operations and functions.  Any change to any of those factors may cause the results to vary.  You should consult other information and performance tests to assist you in fully evaluating your contemplated purchases, including the performance of that product when combined with other products. For more information go to \url{http://www.intel.com/performance}.}

\subsection{Single node, multi-threaded performance}

\begin{figure} [h]
\centering
\includegraphics[width=2.5in]{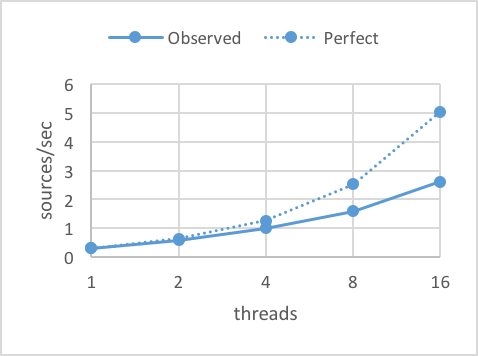}
\caption{Multi-threaded performance. Strong-scaling Celeste on 154 light
sources over up to 16 threads on a Cori Phase~I node. Observe that
scalability drops off beyond 4 threads; this is due to serial garbage
collection.}
\label{thread_scaling}
\end{figure}

A Celeste process uses multiple threads to process multiple light sources
in parallel. We partition runtime as previously described, and show the
number of sources optimized per second at different thread counts in 
Figure~\ref{thread_scaling}. Note the drop-off in scalability beyond 4
threads; this is caused by the serial operation of Julia's garbage collector,
requiring threads to synchronize each GC cycle. Thus,
Amdahl's Law~\cite{hill2008amdahl} limits multi-threaded scalability.
We discuss this challenge 
further in Section~\ref{threads_and_gc}.

To limit time in GC, we use 4 threads per process on our multi-node
runs. A single Cori Phase~I node has 32 cores; we run 8 processes
per node.

\subsection{Weak scaling}

\begin{figure*}[ht]
\includegraphics[width=\linewidth]{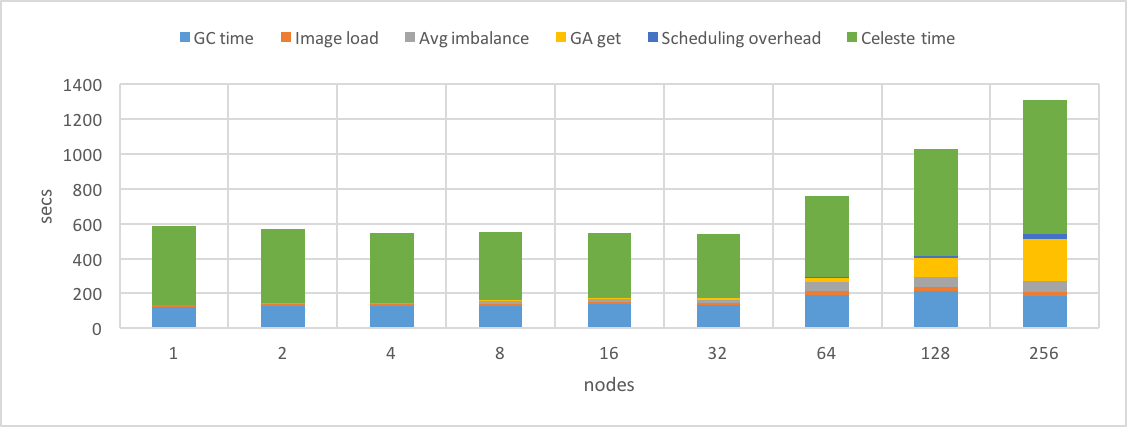}
\caption{Weak-scaling Celeste. While the total runtime shown is precise,
the components of runtime shown are averaged across nodes and thus
should be treated as being representative rather than exact.}
\label{weak_scaling}
\end{figure*}

Figure~\ref{weak_scaling} shows the components of runtime for our weak scaling
experiment. Garbage collection takes between 15\% and 25\% of runtime at
all scales. Image load time does not exceed 1\% of runtime.

Load imbalance is 6.5\% of runtime at most. In absolute terms, 256 nodes
exhibits the highest average load imbalance, yet the runtime attributed
to load imbalance is less than 70 seconds. This is an acceptable level of
imbalance: some individual tasks alone take more than double this time.

The portion of runtime spent retrieving data from the global array is 
the most concerning. Retrieval time is negligible at 64 nodes and fewer, but grows
to 18\% of runtime at 256 nodes. Such a growth rate indicates that transferring
images between nodes is saturating the fabric bandwidth.

Recall from Section~\ref{parallelism}, that Celeste faces a trade-off
between I/O burden and load balance. The use of global arrays is intended
to reduce the I/O burden, but although the network is far faster than
disk, there are nonetheless bandwidth limitations.

Each image is roughly 120 MB in size, and the average task execution time
is less than 5 seconds. If every thread in every process executes a task
that requires a different image, then every 5 seconds on average, a node
will fetch 3.75 GB of image data from across the fabric. Clearly, as the
node count increases, even a high performance fabric such as Cray's Aries
HSN would be overwhelmed---for this worst case at 256 nodes, 192 GB of
images would have to move across the fabric every second.

These results indicate that we have not sufficiently reduced the I/O
burden, however further reduction in I/O will cause increased load
imbalance. We will pursue more advanced strategies in future work.

We note the slight increase in scheduling overhead on the 256 node run;
this is due to global arrays traffic saturating the interconnect and 
slowing down the task scheduler, rather than to any issue with the scheduler itself.

Figure~\ref{fig:weak_sps} shows the light-sources-per-second rate achieved
at different scales in the weak-scaling experiments. We observe perfect
scaling up to 64 nodes followed by a degradation primarily due to slower
fetches from the images global array.

\subsection{Strong scaling}

\begin{figure*}[ht]
\includegraphics[width=\linewidth]{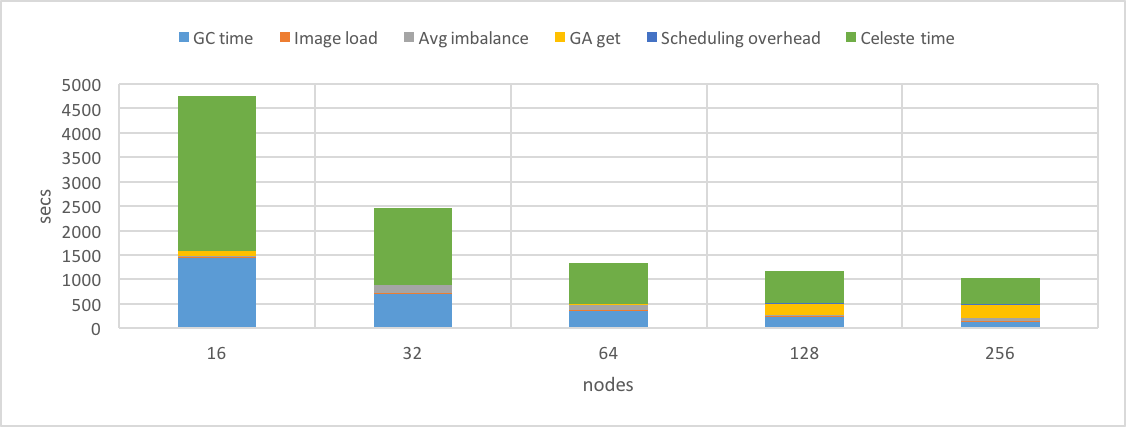}
\caption{Strong-scaling Celeste. Note the reduction in GC time
correlates with reduction in runtime, whereas the increase in
global arrays fetch time correlates with the increase in nodes.}
\label{strong_scaling}
\end{figure*}

\begin{figure*}
\centering
\begin{subfigure}[b]{0.4\textwidth}
\includegraphics[width=\textwidth]{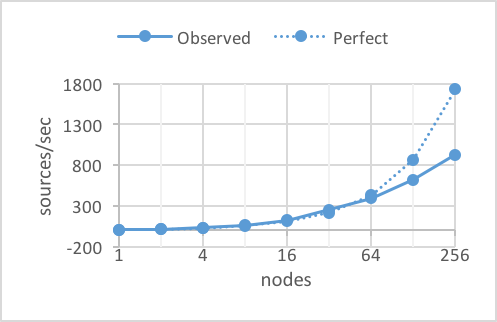}
\caption{Weak scaling}
\label{fig:weak_sps}
\end{subfigure}
\begin{subfigure}[b]{0.4\textwidth}
\includegraphics[width=\textwidth]{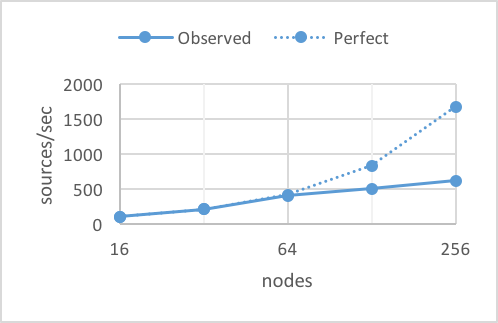}
\caption{Strong scaling}
\label{fig:strong_sps}
\end{subfigure}
\caption{Celeste sources/second. We observe perfect scaling up 
to 64 nodes, after which we are limited by interconnect bandwidth.}
\end{figure*}

Our strong scaling experiments process a region of the sky containing
332,631 light sources. Figure~\ref{strong_scaling} shows the runtimes,
and their breakdowns by routine, at each scale.

Garbage collection time is highest (30\% of total runtime) for the 16
node run, the run with the fewest nodes. That indicates that Julia's serial GC is
detrimental not only for large numbers of threads, but also for long running
processes. For the 256 node strong-scaling run, garbage collection time is
11\% of total runtime.

This finding further underscores the need for improved GC; see
Section~\ref{threads_and_gc} for further discussion.

The next largest portion of total runtime, global arrays retrieval time, does not
exceed 2\% of runtime at 16 nodes. However, at 256 nodes, we find that
it reaches 26\% of runtime, reaffirming the need for improvements in
task scheduling to reduce the I/O burden.

The light-sources-per-second rates achieved at different scales in the 
strong-scaling experiment are shown in Figure~\ref{fig:strong_sps}. Again,
we see perfect scaling up to 64 nodes, followed by a drop-off due to
serial GC and fabric bandwidth saturation.

%% file: scientific_results.tex
\section{Discussion: Scientific results}
For astronomical images, ground truth is unknowable.  That in part is what
makes Bayesian models of the data desirable for astronomy: if their output
follows a plausible model of the data, then it too is plausible. Nonetheless,
validation is essential: every model entails assumptions about the process
that generated the data.

A particular region of the sky, known as Stripe 82, has been imaged everywhere at least
30 times. Most other regions have been imaged just once.  With images from so
many exposures, galaxies and stars that would be faint and hard to accurately
characterize with just one exposure are easily resolved.

``Photo''~\cite{lupton2005sdss} is the current state-of-the-art software pipeline for
constructing large astronomical catalogs.  Photo is a carefully hand-tuned
heuristic. Running Photo on all 30+ exposures of Stripe 82 generates a catalog
that we let stand in for ground truth in our subsequent analysis.

The first numeric column of Table~\ref{scientific_results} shows average error
for Photo itself fit to just one segment of one exposure from Stripe 82.  The
second numeric column shows average error for Celeste fit to the same data.
Celeste has lower error than on Photo by nearly 30\% for locating stars and
galaxies---an improvement that is both statistically significant and of
practical significance to astronomers. For all four colors, Celeste reduces
the error rate by large amounts---always at least 30\%.

For brightness, Photo outperforms Celeste significantly.  That may be due to
systematic errors by Photo, reflected in both ground truth as well as Photo's
predictions based on one exposure, but we do not have firm conclusions yet.  We
have seen cases where a light source is over-saturated, that is, it is too
bright for all the photons from it be counted, yet not flagged as being
over-saturated. In these cases the ground truth does not reflect reality, but
Photo run on all the data largely agrees with itself run on one exposure.

Photo is more likely to misclassify stars as galaxies whereas Celeste is more
likely to misclassify galaxies as stars.  Photo is more accurate at
determining the scales (sizes) of galaxies whereas Celeste is more accurate at
determining their eccentricities and angles.

We anticipate that adjustments to the Celeste model, as discussed in 
Section~\ref{conclusion}, will enable Celeste to outperform Photo across
board. Already, Celeste's results are the new state-of-the-art
by a wide margin for location and color, and thus should be released.

\begin{table}
\caption{Average error on celestial bodies from Stripe 82.}

\centering
\renewcommand{\arraystretch}{1.3}
\begin{tabular}{|l|r|r|}

\hline
%\cline{2-3}
\rule{0pt}{8pt}        & Photo & Celeste \\

\hline
\rule{0pt}{9pt}position     & 0.33     & \textbf{0.24}\\
missed gals  & 0.06 & \textbf{0.03}\\
missed stars & \textbf{0.02}  & 0.09\\
brightness   & \textbf{0.22}     & 0.37 \\
color u-g    & 1.23     & \textbf{0.69} \\
color g-r    & 0.40     & \textbf{0.22} \\
color r-i    & 0.26     & \textbf{0.17} \\
color i-z    & 0.30     & \textbf{0.13} \\
profile      & 0.28     & 0.29 \\
eccentricity & 0.19     & \textbf{0.13} \\
scale        & \textbf{1.97}     & 2.79 \\
angle        & 17.60     & \textbf{12.91} \\

\hline
\end{tabular}

\vspace{10px}
\begin{flushleft}
  	\textbf{Lower is better.} Results in bold are better by more than 2 standard deviations.
    ``Position'' is error, in pixels, for the location of the celestial bodies'
    centers.
    ``Missed gals'' is the proportion of galaxies labeled as stars.
    ``Missed stars'' is the proportion of stars labeled as galaxies.
    ``Brightness'' measures the reference band (r-band) magnitude.
    ``Colors'' are ratios of magnitudes in consecutive bands.
    ``Profile'' is a proportion indicating whether a galaxy is de Vaucouleurs
    or exponential.
    ``Eccentricity'' is the ratio between the lengths of a galaxy's minor and
    major axes.
    ``Scale'' is the effective radius of a galaxy in arcseconds.
    ``Angle'' is the orientation of a galaxy in degrees.
\end{flushleft}

\label{scientific_results}
\end{table}

%% file: julia.tex
\section{Discussion: Julia for HPC and Big Data Analytics}\label{julia}

Developing Celeste in Julia has generally been a positive experience:
rapid prototyping, the high level of abstraction, and a rich set of
libraries have allowed for high productivity. Julia's compiler has
produced high performance code. Our greatest challenge arose from our
use of Julia's threads, a feature clearly identified as experimental.

\subsection{Threads and garbage collection}\label{threads_and_gc}

Celeste uses Julia's experimental multi-threading capability for shared
memory parallelism. While this capability is reasonably stable and allows
excellent scaling, we discovered an important limitation: Julia's garbage
collector currently does not scale well. With 16 threads per process, 
the time consumed by garbage collection exceeds a third of total runtime.
The GC time approaches half of total runtime for longer duration jobs.

The foremost reason for excessive GC time is that Julia's garbage collector is serial.
Thus, Amdahl's Law limits scaling. In addition, serial GC requires
synchronization of all threads before a collection cycle. Celeste's work
is irregular and threads run mostly asynchronously, which adds additional
overhead as compared to applications with regular work such as PDE solvers.

Parallel garbage collection is not a solved problem; there is active
research in this area~\cite{jones2011gc}. However, there are a number of
well-understood methods to improve scalability, such as parallelizing the
marking phase and doing sweeps concurrently with running threads. There
are efforts underway to improve Julia's memory management in these and
other ways. Currently, however, to scale beyond 4 to 6 threads requires
the programmer to carefully consider memory utilization---avoiding copies
and temporaries, and extensively using in-place operations---which
hinders the ease and expressiveness of programming in Julia.

Additionally, while Julia's runtime is thread-safe, much of the standard
library currently is not. That too limits the ease of writing multi-threaded
programs in Julia.

%% file: conclusions.tex
\section{Conclusions and Future Work}\label{conclusion}
We developed a multi-threaded, distributed version of Celeste, and 
demonstrated good weak and strong scaling on up to 256 nodes.
Using this version of Celeste, we generated a catalog for a subset of
SDSS that contains better 
parameter estimates for location, color, galaxy eccentricity, and galaxy angle.
This new catalog is the largest ever to include uncertainty estimates that
follow from a realistic model of the data.

In the process of generating the catalog, we scaled Bayesian inference 
to datasets larger than those reported in the literature, through a
combination of mathematical and algorithmic innovations. A widely held view
is that Bayesian inference works for small datasets, when accuracy is 
paramount, but that heuristics are typically necessary.  Our results suggest
otherwise: given a realistic model, the posterior can be approximated within 
the available computational budget. Celeste is a unique project that has 
advanced the frontier of scalable inference methods in the context of a truly
large-scale, important scientific application on an advanced HPC architecture. 

We have several methodological innovations planned for future work.  First, we
aim to find a stationary point of our objective function by optimizing all
light sources jointly, rather than optimizing each light source with all other
light sources' parameters fixed to their estimates from previous surveys.  That
requires greater communication among nodes.

Second, we plan to allow candidate light sources to ``turn off'', effectively
estimating the true number of light sources from an oversubscribed list of
candidates.

Third, we will apply linear response methods~\cite{giordano2015linear}
to improve the uncertainty quantification of our variational inference
procedure. These methods should mitigate bias stemming from our
choice of a variational distribution that factorizes and provide us with 
uncertain estimates for quantities that we model as unknown constants rather than random variables.

%though we focused on the accuracy of posterior means, Bayesian inference is also attractive because it provide measures of uncertainty.   Unfortunately, variational approximations such as ours are known to under-estimate posterior variance, and parameters that are optimized but not modeled do not have uncertainty estimates at all.  We plan to apply recent linear response methods [CITE Linear response methods for accurate covariance estimates from mean field variational Bayes] to our model to recover accurate posterior covariances for all parameters.

Fourth, we anticipate switching from deterministic to stochastic optimization.
By basing updates on samples from the variational distribution, stochastic
optimization can approximate the posterior for arbitrary combinations of models
and classes of candidate variational distributions, whereas deterministic
optimization is limited to combinations that lead to an analytic objective
function.  That frees us to experiment with even more realistic models of the
data, and even more expressive distributions for approximating the posterior.

Stochastic optimization, however, is likely to be orders of magnitude more
computationally expensive than our current highly optimized form of
second-order deterministic optimization.  Fortunately, the upcoming 
integration of Cori Phase II (based on Intel Phi Knights Landing processors)
will increase the system's computational throughput by roughly one order of magnitude.  An additional order of magnitude speedup may follow from subsampling the pixels,
in addition to sampling from the variational distribution---an approach known
as ``doubly'' stochastic VI~\cite{titsias2014doubly}.  Another order of magnitude 
speedup may follow from using a second-order stochastic
method~\cite{blanchet2016convergence} rather than stochastic gradient descent.

Finally, applying Celeste to multiple images surveys jointly is a promising direction. In conjunction with SDSS, we will process the Dark Energy Camera Legacy Survey (DECaLS)~\cite{decals} in future work.